# Analysis of accretion disk around a black hole in dRGT massive gravity

Sobhan Kazempour[1,a], Yuan-Chuan Zou[2,b], Amin Rezaei Akbarieh[1,c]

[1] Faculty of Physics, University of Tabriz, Tabriz 51666-16471, Iran
[2] School of Physics, Huazhong University of Science and Technology, Wuhan 430074, China

**Abstract** We show the analysis of a thin accretion disk around a static and spherically symmetric black hole in dRGT massive gravity. We present the accretion disk analysis in a gravitational theory with a nonzero graviton mass. Also, we study the event horizons of the black hole and we calculate the equations of motion and effective potential. In the following, we obtain the specific energy, specific angular momentum, and angular velocity of the particles which move in circular orbits. In addition, we plot the effective potentials for two cases and we show the locations of stable circular orbits. At last, we show the possibility of constraining the parameter space of dRGT massive gravity by the astrophysical gamma-ray bursts.

## 1 Introduction

Several observational evidences such as supernovas Ia [1,2], CMB [3,4] and baryon acoustic oscillations [5,6] have indicated that there exist accelerated expansion of the Universe. It is obvious that general relativity can not explain the origin of the current accelerated expansion of the Universe [7,8]. Therefore, it is noticeable that massive gravity theory can be considered as a modification of general relativity for describing the current accelerated expansion of the Universe without a dark energy component [9].

The massive gravity theory was introduced by Fierz and Pauli in 1939 [10]. They introduced the unique Lorentz-invariant linear theory without ghosts by providing consistent interaction terms which are interpreted as a graviton mass. Moreover, van Dam, Veltman, and Zakharov found that the Fierz and Pauli theory suffered from discontinuity in predictions in the limit of $m \to 0$, which is called van Dam-Veltman-Zakharov (vDVZ) discontinuity [11,12]. This way, Vainshtein found that in order to avoid the vDVZ discontinuity, the non-linear massive gravity should be considered and it can be used for recovering the predictions made by general relativity [13]. However, Boulware and Dieser claimed that the non-linear theory has a ghost instability which is called the Boulware-Dieser ghost [14]. Also, this issue was admitted by Hamed Arkani et al. and Creminelli et al. [15,16]. Therefore, in 2010 de Rham, Gabadadze, and Tolley noted that it can be possible to have a ghost-free non-linear massive gravity theory [17].

It should be mentioned that the black holes are the ideal objects for studying the modification theories of general relativity [18]. The significance of this issue lies in the fact that there are differences between the predictions of general relativity and alternative theories. Some studies have been done for finding the spherically symmetric black holes in various massive gravity theories [19–28]. In particular, the spherically symmetric solutions in de Rham–Gabadadze–Tolley (dRGT) massive gravity were obtained in [24,25]. In addition, the charged black hole solution in dRGT was done by [26], and recently the gravitational wave echoes from black holes in massive gravity have been studied by [29].

In this stage, it is interesting to note that the study of accretion disks around compact objects is one of the possible methods for showing the difference between general relativity and alternative theories. We know that the mass of black holes can grow by the accretion disks around them which means that there exist gas clouds as an accretion disk [30]. Note that several studies which are related to the accretion disks have been done throughout the years. In this study [31], the mass of the accretion disk around rotating black holes was investigated. Moreover, the radiation properties of the thin accretion disks and the general relativistic properties of thin accretion disks have been studied in these Refs. [31,32]. Meanwhile, the physical properties and characteristics of matter forming thin accretion disks in static and spherically symmetric wormhole spacetime have been investigated in [33]. In this research [34], the state of the properties of the electromagnetic radiation emitted from the Kerr black hole has been reviewed. The optical appearance of a thin accretion disk around compact objects in the context of the Einstein–Gauss–Bonnet gravity has been considered in this research [35]. The other studies can be found in the Refs. [36–41]

[a] e-mail: s.kazempour@tabrizu.ac.ir (corresponding author)
[b] e-mail: zouyc@hust.edu.cn
[c] e-mail: am.rezaei@tabrizu.ac.ir

This way, we look at different aspects of this issue. Gamma ray bursts (GRBs) are outbursts of huge energy in order of 1% of the solar mass, while this energy only lasts seconds [42,43]. The central engine is generally believed to be an accretion disk around a newly formed black hole. The black hole accretion disk system powers a pair of highly relativistic speeding jets, which produces the GRB if the jet is pointing to the observer. To power so much energy in such short timescales, as well as to power the highly relativistic jet, requires the high efficiency of the gravitational energy of the progenitor star or binary. Consequently, the successful launching of a pair of GRB jets can be used to test the parameter space of the dRGT massive gravity.

The goal of this paper is analysing the accretion disk around the black hole in dRGT massive gravity theory. We indicate the parameters of the accretion disk of the black hole and the possibility of GRB emission. The outline of this paper is organized as follows. In Sect. 2 we present and review the non-linear dRGT massive gravity theory and a static and spherically symmetric black hole solution in this theory. In addition, we perform the analysis of the horizon of this black hole. In the following, we show the calculation of equations of motion and effective potential. In Sect. 3 we calculate all parameters of the accretion disk of the black hole in dRGT massive gravity. Moreover, we demonstrate the numerical analysis which consists of the location of stable circular orbits for two cases. Also, we discuss the emission of the GRB from the accretion disk of the black hole in dRGT massive gravity. Finally, in Sect. 4 we conclude with a discussion. There have been used units which fix the speed of light and the gravitational constant (i.e. $8\pi G = c^4 = 1$).

## 2 Space-time

In this stage, we review the non-linear dRGT massive gravity theory. It should be mentioned that this theory is free from BD ghost at the fully non-linear level and it can be described by a physical metric $g_{\mu\nu}$ and four scalar fields which called stuckelberg fields, $\phi^a (a = 0, 1, 2, 3)$. The action includes the Ricci scalar $R$, a dynamical metric $g_{\mu\nu}$ and its determinant $\sqrt{-g}$. This theory consists of Einstein-Hilbert action and non-linear interaction terms as follows [17,44]

$$S = \int d^4x \sqrt{-g} \frac{1}{2k^2} \left[ R + m_g^2 U(g, \phi^a) \right], \quad (1)$$

where the potential $U(g, \phi^a)$ is the part of the action that provides the mass to the graviton and consists of three parts

$$U(g, \phi^a) = U_2 + \alpha_3 U_3 + \alpha_4 U_4, \quad (2)$$

here $\alpha_3$ and $\alpha_4$ are dimensionless free parameters of the theory. $U_i (i = 2, 3, 4)$ is given by,

$$\begin{aligned} U_2 &= [\mathcal{K}]^2 - [\mathcal{K}^2] \\ U_3 &= [\mathcal{K}]^3 - 3[\mathcal{K}][\mathcal{K}^2] + 2[\mathcal{K}^3], \\ U_4 &= [\mathcal{K}]^4 - 6[\mathcal{K}]^2[\mathcal{K}^2] + 8[\mathcal{K}][\mathcal{K}^3] + 3[\mathcal{K}^2]^2 \\ &\quad - 6[\mathcal{K}^4], \end{aligned} \quad (3)$$

we know that the building block tensor $\mathcal{K}$ is defined as

$$\mathcal{K}^\mu_\nu = \delta^\mu_\nu - \sqrt{g^{\mu\sigma} f_{ab} \partial_\sigma \phi^a \partial_\nu \phi^b}, \quad (4)$$

where the square brackets denote a trace and $f_{ab}$ is the fiducial metric.

As it has been mentioned in [20], new parameters of $\alpha$ and $\beta$ are introduced by dimensionless free parameters of the theory $\alpha_3$ and $\alpha_4$.

$$\alpha_3 = \frac{\alpha - 1}{3}, \qquad \alpha_4 = \frac{\beta}{4} + \frac{1-\alpha}{12}. \quad (5)$$

Note that there are some problems in the class of the black hole solutions in dRGT massive gravity such as superluminality, the Cauchy problem, and strong coupling [45–49].

It is worth mentioning that there exist two types of black hole solutions in dRGT massive gravity. On the first one, the dynamical and fiducial metrics are not simultaneously diagonal which shows no Yukawa suppression at large distances. Thus, it should be encountered with strongly coupled. On the second one, the dynamical and fiducial metrics are simultaneously diagonal. These solutions demonstrate coordinate-invariant singularities at the horizon [50].

To avoid these problems, we follow another approach. We can consider an appropriate form to simplify the calculation. In this paper, we follow [19–23] by considering the fiducial metric as follows

$$f_{\mu\nu} = diag(0, 0, c^2, c^2 sin^2\theta), \quad (6)$$

here $c$ is a constant.

We consider a static and spherically symmetric black hole solution in dRGT massive gravity theory [20].

$$ds^2 = -f(r)dt^2 + \frac{dr^2}{f(r)} + r^2(d\theta^2 + sin^2\theta d\varphi^2), \quad (7)$$

where

$$f(r) = 1 - \frac{2M}{r} + \frac{\Lambda}{3}r^2 + \gamma r + \varsigma, \quad (8)$$

$$\Lambda = 3m_g^2(1 + \alpha + \beta),$$
$$\gamma = -cm_g^2(1 + 2\alpha + 3\beta),$$
$$\varsigma = c^2 m_g^2(\alpha + 3\beta). \quad (9)$$

It should be pointed out that $M$ is an integration constant and, related to the mass of the black hole. Also, the $\Lambda$ is cosmological constant and the $m_g$ is a graviton mass. It is worth mentioning that the graviton mass can be considered as the cosmological constant in the self-expanding cosmological solution in massive gravity.

We should pay attention that if we consider $m_g = 0$, we have the Schwarzschild black hole. Furthermore, in the case of $c = 0$ (i.e. $\gamma = \varsigma = 0$), if we have $(1 + \alpha + \beta) < 0$, the solution is in the form of Schwarzschild-de Sitter. On the other hand, if we have $(1 + \alpha + \beta) > 0$, the solution is Schwarzschild-Anti-de Sitter. Note that $\varsigma$ is the constant potential which is related to the global monopole solution [51].

### 2.1 Horizons

For analyzing the accretion disk around the black hole in dRGT massive gravity theory we analyze the horizons of this black hole by considering $g^{rr} = 0$ as follows

$$f(r) = 1 - \frac{2M}{r} + \frac{\Lambda}{3}r^2 + \gamma r + \varsigma = 0. \quad (10)$$

Note that in Fig. 1, the number of event horizons can be found; by the number of zeros of each curve which means that the curve cross over from the r axis, and the number of crossing shows the number of zeros. As it is obvious that there are three curves with the colors green, blue, and red for three cases, respectively. In the case of the green curve, the parameters are considered as $M = 1, m_g = 1, c = 1, \alpha = 10$ and $\beta = 0.5$, which is related to event horizons of the black hole in dRGT massive gravity. In the second case which is the blue curve, the parameters are considered as $M = 1, m_g = 1, c = 1, \alpha = -3$ and $\beta = 2.1$, this case is completely similar to the first case, but, it is only shifted in comparison with the first case. Both cases show the existence of three event horizons of the back hole in dRGT massive gravity which is admitted in [20]. However, it is interesting to note that the red curve corresponds to the case $m_g = 0$ which means that the graviton mass is zero. This case recovers the Schwarzschild black hole that has only one event horizon and it has a zero in $r = 2$ which is the Schwarzschild radius.

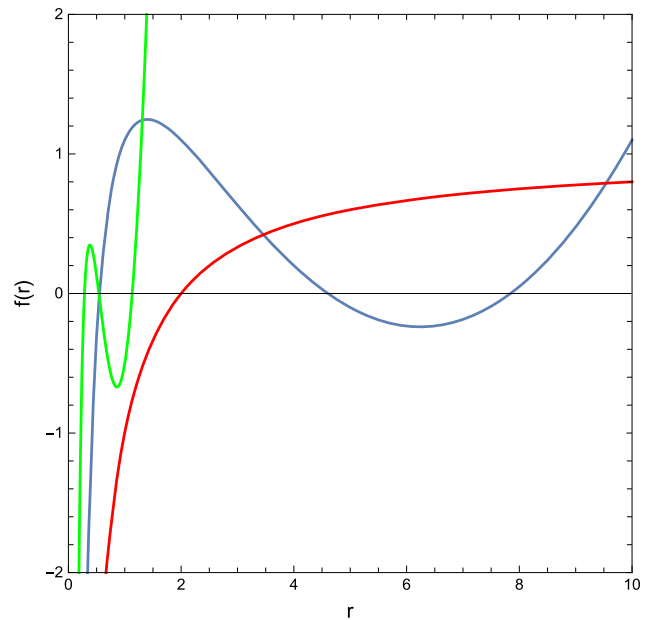

**Fig. 1** The graph illustrates the profile of $f(r)$. The green and blue curves show three event horizons of the black hole in dRGT massive gravity. Also, the red curve is related to the Schwarzschild black hole

### 2.2 Equations of motion

In this step, we calculate the equations of motion and effective potential for evaluating the dynamics of the system. Therefore, we consider the Lagrangian $\mathcal{L}$ for a point particle around the black hole in dRGT massive gravity and this metric is given by Eq. (7). The Lagrangian is

$$\mathcal{L} = \frac{1}{2} g_{\mu\nu} \frac{dx^\mu}{ds} \frac{dx^\nu}{ds} = \frac{1}{2}\varepsilon. \quad (11)$$

It should be mentioned that $\varepsilon = 1$ represents the massive particle and $\varepsilon = 0$ should be considered for the photon. We should consider this issue in the equatorial plane, $E$ conserved energy and $L$ angular momentum can be calculated as follows

$$E = g_{tt}\frac{dt}{ds} = \left(1 - \frac{2M}{r} + \frac{\Lambda}{3}r^2 + \gamma r + \varsigma\right)\frac{dt}{ds}, \quad (12)$$

$$L = g_{\varphi\varphi}\frac{d\varphi}{ds} = r^2 \frac{d\varphi}{ds}. \quad (13)$$

Here, the geodesic equations of a massive particle can be calculated

$$\left(\frac{dr}{ds}\right)^2 = E^2 - \left(1 - \frac{2M}{r} + \frac{\Lambda}{3}r^2 + \gamma r + \varsigma\right)\left(1 + \frac{L^2}{r^2}\right), \quad (14)$$

$$\left(\frac{dr}{d\varphi}\right)^2 = \frac{r^4}{L^2}\left\{E^2 - \left(1 - \frac{2M}{r} + \frac{\Lambda}{3}r^2 + \gamma r + \varsigma\right)\right.$$
$$\left. \times \left(1 + \frac{L^2}{r^2}\right)\right\}, \quad (15)$$

$$\left(\frac{dr}{dt}\right)^2 = \frac{1}{E^2}\left(1 - \frac{2M}{r} + \frac{\Lambda}{3}r^2 + \gamma r + \varsigma\right)^2$$
$$\times \left\{E^2 - \left(1 - \frac{2M}{r} + \frac{\Lambda}{3}r^2 + \gamma r + \varsigma\right)\right.$$
$$\left.\times \left(1 + \frac{L^2}{r^2}\right)\right\}. \tag{16}$$

It is interesting to note that using Eqs. (14–16), we can describe the dynamics of the system completely. Moreover, the effective potential can be obtained by Eq. (14) as follows

$$V_{eff} = \left(1 - \frac{2M}{r} + \frac{\Lambda}{3}r^2 + \gamma r + \varsigma\right)\left(1 + \frac{L^2}{r^2}\right). \tag{17}$$

Here, by considering Eq. (19) we have

$$V_{eff} = \left(1 - \frac{2}{\bar{r}} + \bar{\Lambda}\bar{r}^2 + \bar{\gamma}\bar{r} + \varsigma\right)\left(1 + \frac{\bar{L}^2}{\bar{r}^2}\right), \tag{18}$$

where, there are the dimensionless quantities.

$$\bar{r} = \frac{r}{M}, \qquad \bar{\Lambda} = \frac{\Lambda}{3M^2},$$
$$\bar{\gamma} = \gamma M, \qquad \bar{L} = \frac{L}{M}. \tag{19}$$

## 3 Thin accretion disk

In this step, we can calculate all parameters of the accretion disk of the black hole in dRGT massive gravity.

The specific energy $E$, the specific angular momentum $L$, the angular velocity $\Omega$ and the flux of the radiant energy $F$, over the disk of the particles which move in circular orbits, can be calculated for all cases.

Note that the physical properties of the accretion disk follow from certain structure equations which are related to the conservation of the mass, the energy, and the angular momentum. It is noticeable that the kinematic quantities depend on the radius of the orbit and they can be obtained using the general expressions which are introduced by [52,53]. Note that we calculate all parameters using dimensionless quantities which were introduced in Eq. (19).

$$\Omega = \sqrt{-\frac{g_{tt,r}}{g_{\phi\phi,r}}} = \sqrt{\frac{1}{\bar{r}^3} + \frac{\bar{\gamma}}{2\bar{r}} + \bar{\Lambda}}, \tag{20}$$

$$\bar{E} = -\frac{g_{tt}}{\sqrt{-g_{tt} - g_{\phi\phi}\Omega^2}} = \frac{\bar{r}\left(1 + \bar{r}(\bar{\gamma} + \bar{r}\bar{\Lambda}) + \varsigma\right) - 2}{\bar{r}\sqrt{1 - \frac{3}{\bar{r}} + \frac{\bar{r}\bar{\gamma}}{2} + \varsigma}}, \tag{21}$$

$$\bar{L} = \frac{g_{\phi\phi}\Omega}{\sqrt{-g_{tt} - g_{\phi\phi}\Omega^2}} = \frac{\bar{r}^2\Omega}{\sqrt{1 - \frac{3}{\bar{r}} + \frac{\bar{r}\bar{\gamma}}{2} + \varsigma}}. \tag{22}$$

We can consider $V_{eff} = 0$ and $\frac{dV_{eff}}{dr} = 0$ which means that we impose the conditions on the effective potential for the particle to move in circular orbits. Thus, using Eqs. (18) and (22), we have

$$\frac{d^2 V_{eff}}{dr^2} = \frac{1}{\bar{r}^3(\bar{r}(2 + \bar{r}\bar{\gamma} + 2\varsigma) - 6)}\left\{2\bar{r}\left(2(1 + \varsigma)\right.\right.$$
$$\left.+ \bar{r}(\bar{r}^2\bar{\gamma}^2 + 3\bar{\gamma}(\bar{r}^3\bar{\Lambda} + \bar{r}\varsigma + \bar{r} - 4))\right.$$
$$\left.\left.+ 2\bar{r}\bar{\Lambda}(4\bar{r}(1 + \varsigma) - 15)\right) - 24\right\}. \tag{23}$$

Innermost circular orbits occur at the local minimum of the effective potential, thus the dimensionless radius of the innermost stable circular geodesic orbit is obtained from $\frac{d^2 V_{eff}}{dr^2} = 0$.

Here, we know that the flux of the radiant energy over the disk could be obtained from the following relation [31].

$$F(r) = \frac{-\dot{M}_0}{4\pi\sqrt{-g}}\frac{\Omega_{,r}}{(\bar{E} - \Omega\bar{L})^2}\int_{r_{ms}}^{r}(\bar{E} - \Omega\bar{L})\bar{L}_{,r}dr, \tag{24}$$

which should follow Stefan–Boltzmann law when the disk is supposed to be in thermal equilibrium. Here, the $\dot{M}_0$ is the mass accretion rate.

It should be noted that using the conservation laws the flux of the radiant energy over the disk spanning between the ISCO and a certain radial distance can be achieved [31,53].

In Fig. 2, we have shown the angular velocity $\Omega$ with respect of $\bar{r}$. In fact, we have demonstrated the angular velocity in terms of different values of $\bar{\gamma}$ and $\bar{\Lambda}$ which are dimensionless quantities. Meanwhile, it can be seen that the angular velocity increase slightly by increasing the values of $\bar{\gamma}$ and $\bar{\Lambda}$. In fact, the changes of the angular velocity are very little. As it is obvious that in all cases the graph of angular velocity reaches a pick between $\bar{r} = 0.5$ and $\bar{r} = 1$ and, in the following, it is a plateau.

In Fig. 3, the specific energy $\bar{E}$ of the particles of the thin accretion disk to $\bar{r}$ is indicated for different values of $\varsigma, \bar{\gamma}$ and $\bar{\Lambda}$. From an overall perspective, the most striking feature of the Fig. 3 is that the specific energy witnessed three phases. In the first phase, the energy dropped sharply between $\bar{r} = 1$ and $\bar{r} = 1.5$. The second phase is the lowest level of energy in all cases of values which is around $\bar{r} = 1.5$. In the third phase, the specific energy increases sharply.

In Fig. 4, the graphs illustrated the specific angular momentum $\bar{L}$ to $\bar{r}$ for different values of $\varsigma, \bar{\gamma}$ and $\bar{\Lambda}$. We can immediately see that the specific angular momentum rises totally for all cases in little values of $\bar{r}$ which is around $\bar{r} = 1$. Furthermore, by increasing the values of $\varsigma, \bar{\gamma}$ and $\bar{\Lambda}$ the specific angular momentum increase.

Therefore, we have obtained all parameters of the accretion disk of the black hole in dRGT massive gravity. Moreover, we have analyzed the changes of those parameters in terms of the different values of the components of the theory.

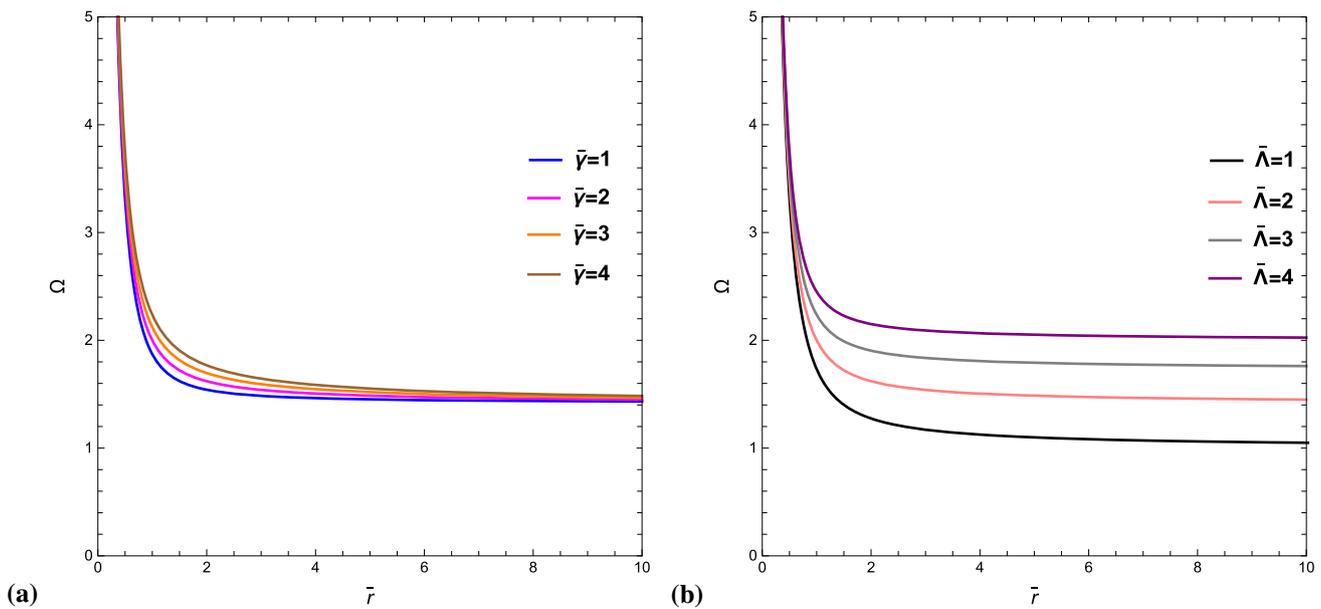

**Fig. 2** The graphs show the angular velocity $\Omega$ to $\bar{r}$ for the particles moving around the black hole and make the thin accretion disk. **a** Different values of $\bar{\gamma}$ with considering $\bar{\Lambda} = 2$. **b** Different values of $\bar{\Lambda}$ with considering $\bar{\gamma} = 2$

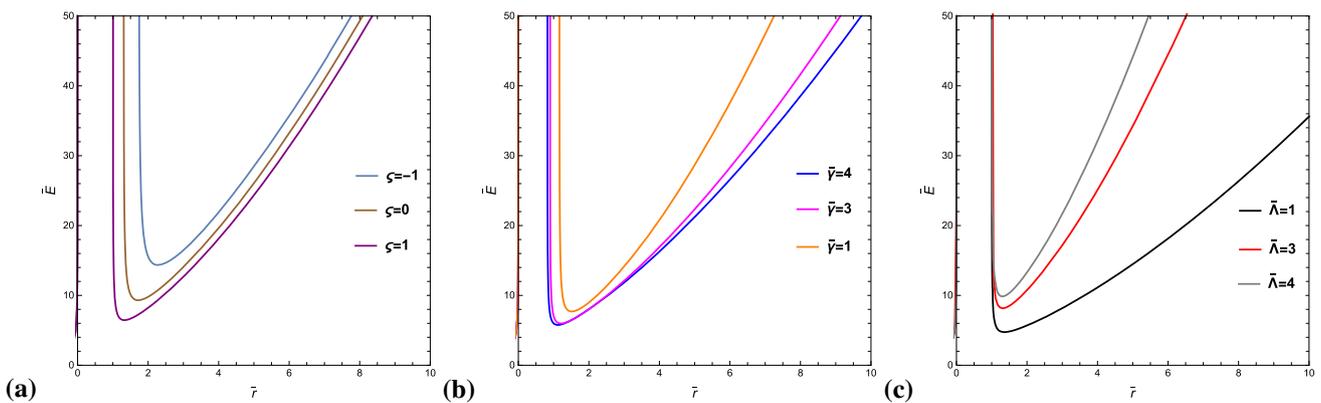

**Fig. 3** The graphs show the specific energy $\bar{E}$ to $\bar{r}$ for the particles moving around the black hole and making the thin accretion disk. **a** Different values of $\varsigma$ with considering $\bar{\gamma} = 2$ and $\bar{\Lambda} = 2$. **b** Different values of $\bar{\gamma}$ with considering $\varsigma = 1$ and $\bar{\Lambda} = 2$. **c** Different values of $\bar{\Lambda}$ with considering $\varsigma = 1$ and $\bar{\gamma} = 2$

### 3.1 Numerical analysis

In this stage, we consider the two cases that were used by Ghosh and et al. [20], also we have used those values in Sect. 2.1. In the next subsection, we will discuss the emission of the gamma-ray burst for these cases. Therefore, we plot the effective potentials and the location of stable circular orbits for these cases.

- In the first case, we consider $M = 1, m_g = 1, c = 1, \alpha = 10$ and $\beta = 0.5$.
- In the second case, we consider $M = 1, m_g = 1, c = 1, \alpha = -3$ and $\beta = 2.1$.

It is worth mentioning that in general relativity a test particle can move around the Schwarzschild black hole and make the smallest marginally stable circular orbit. This orbit is known as an innermost stable circular orbit (ISCO). As ISCO is the inner edge of the disk, it has an essential role in black hole accretion disks [54].

As we have shown in Fig. 1, in the case of $m_g = 0$, we recover the Schwarzschild black hole and the Schwarzschild radius which is $r = 2$ (i.e. red curve in Fig. 1). Meanwhile in Fig. 5, in the case of $m_g = 0$, we have indicated the innermost stable circular orbit for the Schwarzschild black hole which is $r_{ISCO} = 6$. In fact, we have illustrated it by a black dot in Fig. 5. Furthermore, it is possible to have circular orbits between the Schwarzschild radius and ISCO, but they are not stable. In Table 1, we have demonstrated the event horizon and the innermost stable circular orbit for the case of $m_g = 0$ which recovers Schwarzschild black hole.

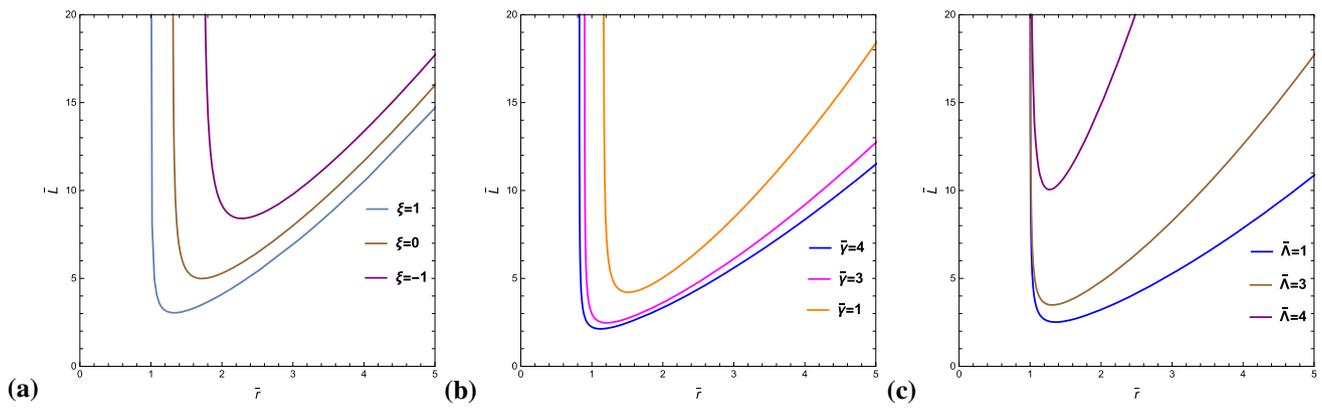

**Fig. 4** The graphs show the specific angular momentum $\bar{L}$ to $\bar{r}$ for the particles moving around the black hole and making the thin accretion disk. **a** Different values of $\varsigma$ with considering $\bar{\gamma} = 2$ and $\bar{\Lambda} = 2$. **b** Different values of $\bar{\gamma}$ with considering $\varsigma = 1$ and $\bar{\Lambda} = 2$. **c** Different values of $\bar{\Lambda}$ with considering $\varsigma = 1$ and $\bar{\gamma} = 2$

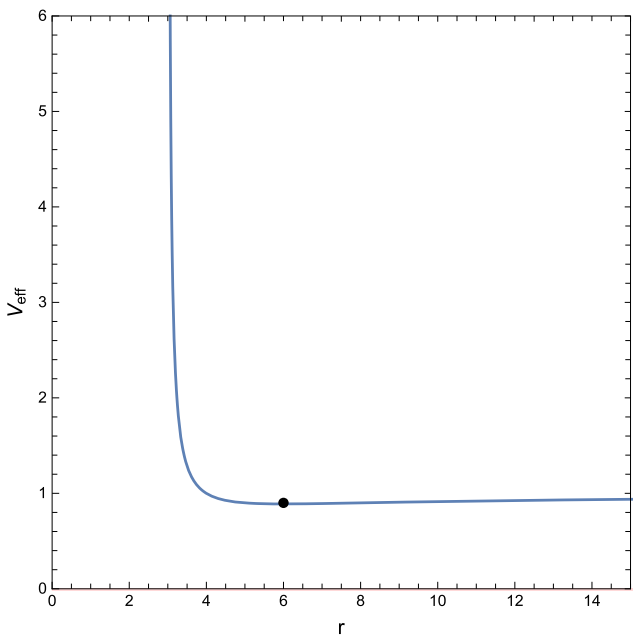

**Fig. 5** Effective potential for the Schwarzschild black hole in the case of $m_g = 0$. The dot indicates the location of the innermost stable circular orbit, $r_{ISCO} = 6$

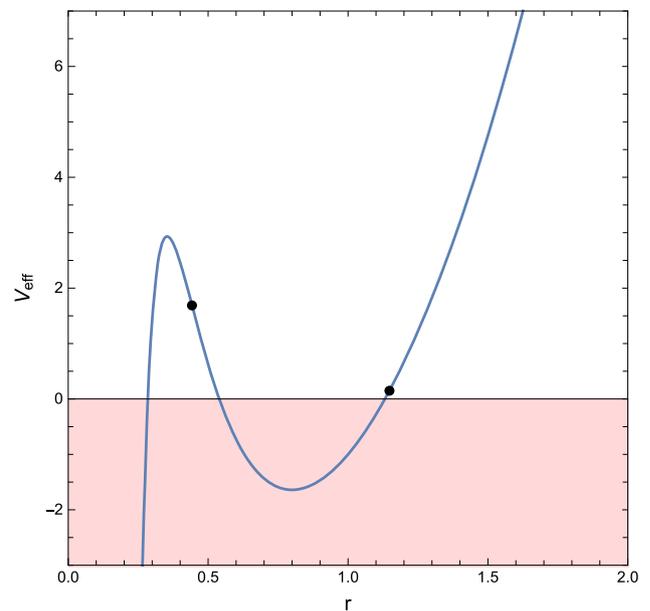

**Fig. 6** Effective potential for the first case with $L = 1$. The dots indicate the location of the stable circular orbits for the black hole in dRGT massive gravity, $r_{SCO} = 0.44$ and $r_{SCO} = 1.15$

**Table 1** Location of the event horizon and the innermost stable circular orbit for the case of ($m_g = 0$) which recovers Schwarzschild black hole

| $m_g$ | Event horizon | Innermost stable circular orbit |
| --- | --- | --- |
| 0 | 2 | 6 |

The effective potentials are plotted using Eq. (17) for both cases which have been mentioned in the first of this subsection. Note that by considering $\frac{d^2 V_{eff}}{dr^2} = 0$, we indicate the locations of the stable circular orbits for the two cases and we show those points using the black dots in Figs. 6 and 7.

It is interesting to note that we have achieved two stable circular orbits for any of those cases. While in the case of $m_g = 0$ there is an innermost stable circular orbit, there are two stable circular orbits for the black hole in dRGT massive gravity. Also, we showed that there are three event horizons around the black hole in dRGT massive gravity. In these two cases, the location of one of the stable circular orbits is always between the inner event horizon and the middle event horizon and the other one is bigger than the outer event horizon. From a realistic perspective, the stable circular orbits which are between the inner event horizon and middle event horizon

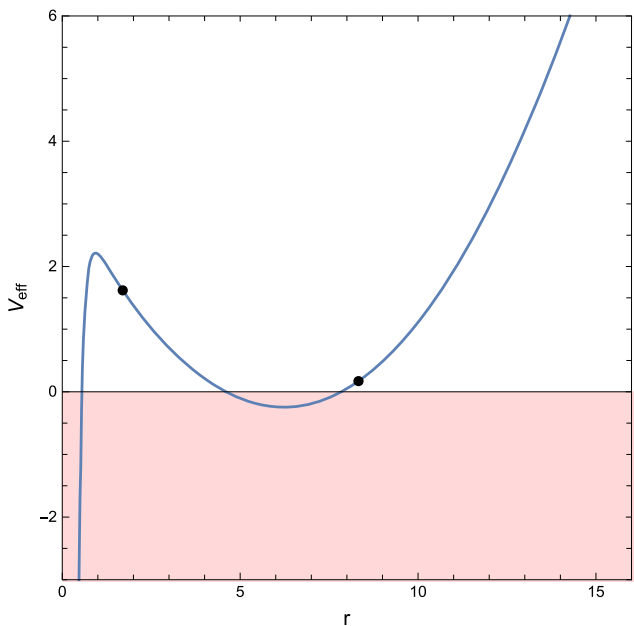

**Fig. 7** Effective potential for the second case with $L = 1$. The dots indicate the location of the stable circular orbits for the black hole in dRGT massive gravity, $r_{SCO} = 1.70$ and $r_{SCO} = 8.33$

can not be observed. However, there are several researchers have admitted the existence of extra event horizons and the bounds orbits among them [55–60]. In the Tables 2 and 3, we have performed the details of event horizons and stable circular orbits in terms of different values of $\beta$ and $\alpha$ for two cases that have been mentioned.

### 3.2 Gamma Ray Burst

In this subsection, the emission of the gamma-ray burst will be discussed for the black hole.

We can consider GRB from the accretion disk of the black hole. In fact, we investigate the possibility of GRB in the accretion disk of the black hole in the dRGT massive gravity.

Notice from Tables 2 and 3, the outer stable circular orbit is always bigger than the outer event horizon. Consequently, the energy that can be extracted from the gravitational energy should be ended at that stable circular orbit. Given an efficiency $\eta$ for converting the gravitational energy to energy of the gamma-rays, one can constrain the parameters of the dRGT massive gravity.

The total energy of a typical gamma-ray burst is $\sim 10^{52}$ erg, the typical mass of the progenitor is $M \sim 10 M_\odot$, the outer stable circular orbit $x$, one gets the released energy

$$E_\gamma = \eta \frac{M}{x}, \qquad (25)$$

suppose $\eta < 0.1$, which gives

$$x < \frac{\eta M}{E_\gamma} \sim 18 \eta_{-1} M_1 E_{\gamma,52}^{-1}, \qquad (26)$$

where the notation is used

$$Q = 10^k \times Q_k, \quad M_1 = M/10 M_\odot. \qquad (27)$$

$M_1$ is a unit of a solar mass. With more realistic observational data, we can put more stringent constraints on the stable circular orbit, and consequently on the allowed parameters of the dRGT massive gravity.

## 4 Conclusion

In this work, we have studied the thin accretion disk around the static and spherically symmetric black hole in dRGT massive gravity. In fact, we would like to get some insights into the differences in dRGT massive gravity in comparison with general relativity. To further studies, we have analyzed the event horizons of the black hole in dRGT massive gravity. Meanwhile, we have presented the equations of motion and effective potential. In the following stage, we have calculated the specific energy, the specific angular momentum, and the angular velocity of the particles which move in circular orbits around the black hole in dRGT massive gravity and we showed the relation of the flux of the radiant energy over the disk. As the necessary condition for the existence of the marginally stable orbit is $\frac{d^2 V_{eff}}{dr^2} = 0$, we have obtained the second derivative of effective potential to radial coordinate. Moreover, we have illustrated the changes of the specific energy, the specific angular momentum, and the angular velocity in terms of the different values of the components of the theory.

Furthermore, we have plotted the effective potentials for two cases and we have indicated the locations of stable circular orbits. Also, we have demonstrated the locations of the event horizons and the stable circular orbits, in the tables for different values of $\beta$ and $\alpha$. Note we have shown that in the case of $m_g = 0$ the Schwarzschild black hole and Schwarzschild radius have been recovered.

Finally, we have discussed the possibility of constraining the parameter space of dRGT massive gravity with observations. With the typical values of gamma-ray bursts, we got the stable circular orbit should be smaller than 18 times the gravitational radius, which is a kind of loose constraint compared with the tables shown in the text. However, with more individual extreme events and more reliable efficiency $\eta$ from simulations, one may get more stringent constraints on $\alpha$ and $\beta$.

**Table 2** Location of the event horizons and the stable circular orbits for the first case ($\alpha = 10$) and different values of $\beta$ around the black hole in dRGT massive gravity

| $\beta$ | Inner event horizon | Middle event horizon | Outer event horizon | Stable circular orbits |
| --- | --- | --- | --- | --- |
| 0.5 | 0.28 | 0.539 | 1.13 | 0.44, 1.15 |
| 0.7 | 0.246 | 0.634 | 1.09 | 0.42, 1.21 |
| 1 | 0.211 | 0.78 | 1 | 0.39, 1.30 |
| 3 | 0.12 | – | – | 0.30, 1.67 |

**Table 3** Location of the event horizons and the stable circular orbits for the second case ($\beta = 2.1$) and different values of $\alpha$ around the black hole in dRGT massive gravity

| $\alpha$ | Inner event horizon | Middle event horizon | Outer event horizon | Stable circular orbits |
| --- | --- | --- | --- | --- |
| $-3.2$ | – | 0.56 | 2.87 | 12.04 |
| $-3$ | 0.55 | 4.60 | 7.84 | 1.70, 8.33 |
| $-2.5$ | 0.534 | – | – | 1.05, 4.97 |
| $-2$ | 0.51 | – | – | 0.96, 3.65 |

**Data Availability Statement** This manuscript has no associated data or the data will not be deposited. [Authors' comment: There is not any data which are related to the paper.]

### References

1. M.M. Phillips, Astrophys. J. Lett. **413**, L105–L108 (1993). https://doi.org/10.1086/186970
2. A.G. Riess et al., Supernova Search Team. Astron. J. **116**, 1009–1038 (1998). https://doi.org/10.1086/300499 arXiv:astro-ph/9805201
3. P.A.R. Ade et al., Planck. Astron. Astrophys. **594**, A13 (2016). https://doi.org/10.1051/0004-6361/201525830 arXiv:1502.01589 [astro-ph.CO]
4. D.N. Spergel et al., WMAP Astrophys. J. Suppl. **148**, 175–194 (2003). https://doi.org/10.1086/377226 arXiv:astro-ph/0302209
5. F. Beutler, C. Blake, M. Colless, D.H. Jones, L. Staveley-Smith, L. Campbell, Q. Parker, W. Saunders, F. Watson, Mon. Not. R. Astron. Soc. **416**, 3017–3032 (2011). https://doi.org/10.1111/j.1365-2966.2011.19250.x arXiv:1106.3366 [astro-ph.CO]
6. W.J. Percival et al., SDSS Mon. Not. R. Astron. Soc. **401**, 2148–2168 (2010). https://doi.org/10.1111/j.1365-2966.2009.15812.x arXiv:0907.1660 [astro-ph.CO]
7. S. Weinberg, Rev. Mod. Phys. **61**, 1–23 (1989). https://doi.org/10.1103/RevModPhys.61.1
8. P.J.E. Peebles, B. Ratra, Rev. Mod. Phys. **75**, 559–606 (2003). https://doi.org/10.1103/RevModPhys.75.559 arXiv:astro-ph/0207347
9. C. de Rham, Living Rev. Relativ. **17**, 7 (2014). https://doi.org/10.12942/lrr-2014-7 arXiv:1401.4173 [hep-th]
10. M. Fierz, W. Pauli, Proc. R. Soc. Lond. A **173**, 211–232 (1939). https://doi.org/10.1098/rspa.1939.0140
11. V.I. Zakharov, JETP Lett. **12**, 312 (1970)
12. H. van Dam, M.J.G. Veltman, Nucl. Phys. B **22**, 397–411 (1970). https://doi.org/10.1016/0550-3213(70)90416-5
13. A.I. Vainshtein, Phys. Lett. B **39**, 393–394 (1972). https://doi.org/10.1016/0370-2693(72)90147-5
14. D.G. Boulware, S. Deser, Phys. Rev. D **6**, 3368–3382 (1972). https://doi.org/10.1103/PhysRevD.6.3368
15. N. Arkani-Hamed, H. Georgi, M.D. Schwartz, Ann. Phys. **305**, 96–118 (2003). https://doi.org/10.1016/S0003-4916(03)00068-X arXiv:hep-th/0210184
16. P. Creminelli, A. Nicolis, M. Papucci, E. Trincherini, JHEP **09**, 003 (2005). https://doi.org/10.1088/1126-6708/2005/09/003 arXiv:hep-th/0505147
17. C. de Rham, G. Gabadadze, Phys. Rev. D **82**, 044020 (2010). https://doi.org/10.1103/PhysRevD.82.044020 arXiv:1007.0443 [hep-th]
18. D. Psaltis, Living Rev. Relativ. **11**, 9 (2008). https://doi.org/10.12942/lrr-2008-9 arXiv:0806.1531 [astro-ph]
19. R.G. Cai, Y.P. Hu, Q.Y. Pan, Y.L. Zhang, Phys. Rev. D **91**(2), 024032 (2015). https://doi.org/10.1103/PhysRevD.91.024032 arXiv:1409.2369 [hep-th]
20. S.G. Ghosh, L. Tannukij, P. Wongjun, Eur. Phys. J. C **76**(3), 119 (2016). https://doi.org/10.1140/epjc/s10052-016-3943-x arXiv:1506.07119 [gr-qc]
21. D. Vegh, arXiv:1301.0537 [hep-th]
22. A. Adams, D.A. Roberts, O. Saremi, Phys. Rev. D **91**(4), 046003 (2015). https://doi.org/10.1103/PhysRevD.91.046003 arXiv:1408.6560 [hep-th]

23. J. Xu, L.M. Cao, Y.P. Hu, Phys. Rev. D **91**(12), 124033 (2015). https://doi.org/10.1103/PhysRevD.91.124033. arXiv:1506.03578 [gr-qc]
24. T.M. Nieuwenhuizen, Phys. Rev. D **84**, 024038 (2011). https://doi.org/10.1103/PhysRevD.84.024038. arXiv:1103.5912 [gr-qc]
25. R. Brito, V. Cardoso, P. Pani, Phys. Rev. D **88**, 064006 (2013). https://doi.org/10.1103/PhysRevD.88.064006. arXiv:1309.0818 [gr-qc]
26. L. Berezhiani, G. Chkareuli, C. de Rham, G. Gabadadze, A.J. Tolley, Phys. Rev. D **85**, 044024 (2012). https://doi.org/10.1103/PhysRevD.85.044024. arXiv:1111.3613 [hep-th]
27. E. Babichev, A. Fabbri, JHEP **07**, 016 (2014). https://doi.org/10.1007/JHEP07(2014)016. arXiv:1405.0581 [gr-qc]
28. E. Babichev, R. Brito, Class. Quantum Gravity **32**, 154001 (2015). https://doi.org/10.1088/0264-9381/32/15/154001. arXiv:1503.07529 [gr-qc]
29. R. Dong, D. Stojkovic, Phys. Rev. D **103**(2), 024058 (2021). https://doi.org/10.1103/PhysRevD.103.024058. arXiv:2011.04032 [gr-qc]
30. C.M. Urry, P. Padovani, Publ. Astron. Soc. Pac. **107**, 803 (1995). https://doi.org/10.1086/133630. arXiv:astro-ph/9506063
31. D.N. Page, K.S. Thorne, Astrophys. J. **191**, 499–506 (1974). https://doi.org/10.1086/152990
32. K.S. Thorne, Astrophys. J. **191**, 507–520 (1974). https://doi.org/10.1086/152991
33. T. Harko, Z. Kovacs, F.S.N. Lobo, Phys. Rev. D **78**, 084005 (2008). https://doi.org/10.1103/PhysRevD.78.084005. arXiv:0808.3306 [gr-qc]
34. C. Bambi, Rev. Mod. Phys. **89**(2), 025001 (2017). https://doi.org/10.1103/RevModPhys.89.025001. arXiv:1509.03884 [gr-qc]
35. G. Gyulchev, P. Nedkova, T. Vetsov, S. Yazadjiev, arXiv:2106.14697 [gr-qc]
36. F.S. Guzman, Phys. Rev. D **73**, 021501 (2006). https://doi.org/10.1103/PhysRevD.73.021501. arXiv:gr-qc/0512081
37. C. Bambi, G. Lukes-Gerakopoulos, Phys. Rev. D **87**(8), 083006 (2013). https://doi.org/10.1103/PhysRevD.87.083006. arXiv:1302.0565 [gr-qc]
38. S. Soroushfar, S. Upadhyay, Eur. Phys. J. Plus **135**(3), 338 (2020). https://doi.org/10.1140/epjp/s13360-020-00329-4. arXiv:2003.08185 [gr-qc]
39. D. Pugliese, Z. Stuchlík, Class. Quantum Gravity **38**(14), 14 (2021). https://doi.org/10.1088/1361-6382/abff97. arXiv:2108.07069 [gr-qc]
40. S. Gimeno-Soler, J.A. Font, C. Herdeiro, E. Radu, arXiv:2106.15425 [gr-qc]
41. D.V. Boneva, E.A. Mikhailov, M.V. Pashentseva, D.D. Sokoloff, Astron. Astrophys. **652**, A38 (2021). https://doi.org/10.1051/0004-6361/202038680. arXiv:2107.02766 [astro-ph.HE]
42. T. Piran, Rev. Mod. Phys. **76**, 1143–1210 (2004). https://doi.org/10.1103/RevModPhys.76.1143. arXiv:astro-ph/0405503 [astro-ph]
43. P. Kumar, B. Zhang, Phys. Rep. **561**, 1–109 (2014). https://doi.org/10.1016/j.physrep.2014.09.008. arXiv:1410.0679 [astro-ph.HE]
44. C. de Rham, G. Gabadadze, A.J. Tolley, Phys. Rev. Lett. **106**, 231101 (2011). https://doi.org/10.1103/PhysRevLett.106.231101. arXiv:1011.1232 [hep-th]
45. F. Capela, P.G. Tinyakov, JHEP **04**, 042 (2011). https://doi.org/10.1007/JHEP04(2011)042. arXiv:1102.0479 [gr-qc]
46. P. Motloch, W. Hu, A. Joyce, H. Motohashi, Phys. Rev. D **92**(4), 044024 (2015). https://doi.org/10.1103/PhysRevD.92.044024. arXiv:1505.03518 [hep-th]
47. S. Deser, K. Izumi, Y.C. Ong, A. Waldron, Phys. Lett. B **726**, 544–548 (2013). https://doi.org/10.1016/j.physletb.2013.09.001. arXiv:1306.5457 [hep-th]
48. S. Deser, K. Izumi, Y.C. Ong, A. Waldron, https://doi.org/10.1142/9789814590112_0029. arXiv:1312.1115 [hep-th]
49. K. Izumi, Y.C. Ong, Class. Quantum Gravity **30**, 184008 (2013). https://doi.org/10.1088/0264-9381/30/18/184008. arXiv:1304.0211 [hep-th]
50. R.A. Rosen, JHEP **10**, 206 (2017). https://doi.org/10.1007/JHEP10(2017)206. arXiv:1702.06543 [hep-th]
51. S.H. Rhie, D.P. Bennett, Phys. Rev. Lett. **67**, 1173 (1991). https://doi.org/10.1103/PhysRevLett.67.1173
52. T. Harko, Z. Kovacs, F.S.N. Lobo, Phys. Rev. D **79**, 064001 (2009). https://doi.org/10.1103/PhysRevD.79.064001. arXiv:0901.3926 [gr-qc]
53. G. Gyulchev, P. Nedkova, T. Vetsov, S. Yazadjiev, Phys. Rev. D **100**(2), 024055 (2019). https://doi.org/10.1103/PhysRevD.100.024055. arXiv:1905.05273 [gr-qc]
54. C.W. Misner, K.S. Thorne, J.A. Wheeler
55. E. Hackmann, V. Kagramanova, J. Kunz, C. Lammerzahl, Phys. Rev. D **78**, 124018 (2008). https://doi.org/10.1103/PhysRevD.78.124018. arXiv:0812.2428 [gr-qc]
56. E. Hackmann, C. Lammerzahl, Phys. Rev. Lett. **100**, 171101 (2008). https://doi.org/10.1103/PhysRevLett.100.171101. arXiv:1505.07955 [gr-qc]
57. S. Grunau, V. Kagramanova, Phys. Rev. D **83**, 044009 (2011). https://doi.org/10.1103/PhysRevD.83.044009. arXiv:1011.5399 [gr-qc]
58. S. Soroushfar, R. Saffari, S. Kazempour, S. Grunau, J. Kunz, Phys. Rev. D **94**(2), 024052 (2016). https://doi.org/10.1103/PhysRevD.94.024052. arXiv:1605.08976 [gr-qc]
59. S. Grunau, B. Khamesra, Phys. Rev. D **87**(12), 124019 (2013). https://doi.org/10.1103/PhysRevD.87.124019. arXiv:1303.6863 [gr-qc]
60. S. Kazempour, S. Soroushfar, Chin. J. Phys. **65**, 579–592 (2020). https://doi.org/10.1016/j.cjph.2020.04.004. arXiv:1709.06541 [gr-qc]